# Oxidising and carburising catalyst conditioning for the controlled growth and transfer of large crystal monolayer hexagonal boron nitride

Vitaliy Babenko[1], Ye Fan[1], Vlad-Petru Veigang-Radulescu[1,2], Barry Brennan[2], Andrew J. Pollard[2], Oliver Burton[1], Jack A. Alexander-Webber[1], Robert S. Weatherup[3], Barbara Canto[4], Martin Otto[4], Daniel Neumaier[4], Stephan Hofmann[1]

[1] Department of Engineering, University of Cambridge, 9 JJ Thomson Avenue, Cambridge, CB3 0FA, United Kingdom
[2] National Physical Laboratory, Hampton Rd, Teddington, Middlesex, TW11 0LW, United Kingdom
[3] Department of Materials, The University of Oxford, Parks Road, Oxford, OX1 3PH, United Kingdom
[4] Advanced Microelectronic Center Aachen (AMICA), AMO GmbH, Otto-Blumenthal-Straße 25, 52074 Aachen, Germany

E-mail: sh315@cam.ac.uk

**Abstract**

Hexagonal boron nitride (h-BN) is well-established as a requisite support, encapsulant and barrier for 2D material technologies, but also recently as an active material for applications ranging from hyperbolic metasurfaces to room temperature single-photon sources. Cost-effective, scalable and high quality growth techniques for h-BN layers are critically required. We utilise widely-available iron foils for the catalytic chemical vapour deposition (CVD) of h-BN and report on the significant role of bulk dissolved species in h-BN CVD, and specifically, the balance between dissolved oxygen and carbon. A simple pre-growth conditioning step of the iron foils enables us to tailor an error-tolerant scalable CVD process to give exceptionally large h-BN monolayer domains. We also develop a facile method for the improved transfer of as-grown h-BN away from the iron surface by means of the controlled humidity oxidation and subsequent rapid etching of a thin interfacial iron oxide; thus, avoiding the impurities from the bulk of the foil. We demonstrate wafer-scale (2") production and utilise this h-BN as a protective layer for graphene towards integrated (opto-)electronic device fabrication.

## 1. Introduction

Hexagonal boron nitride (h-BN) is a well-known member of the 2D family of materials, isostructural to graphene, but with distinct properties. Due to its wide bandgap (~5.9 eV[1]) and chemical inertness, h-BN has become indispensable as a support material and encapsulant for graphene[2,3] and an increasing range of other 2D materials[4], as well as a barrier layer in van der Waals heterostructures[5]. h-BN has also attracted interest as an active material for infrared hyperbolic metasurfaces[6,7] and defect-induced sub-bandgap single photon emission at room temperature[8-11].

A major technological challenge is how to effectively integrate h-BN in scalable process flows for these applications. Significant progress has been made in the chemical vapour deposition (CVD) of h-BN from surface science studies in the nineties[12,13] to targeted production of high-quality h-BN layers on various catalytic transition metals[14-22]. Similar to graphene, such catalytic CVD has become the dominant approach for producing large-area continuous h-BN films and achieving h-BN crystal domain sizes much larger than possible via exfoliation[16,17,22]. For most applications the critical challenges include scalable CVD process development to achieve high quality h-BN material at lowest cost, and to effectively and cleanly transfer the h-BN away from the growth catalyst for device integration. In analogy to graphene, the strength of the h-BN/metal interactions can be used to guide the choice of the catalyst[22], whereby too strong an interaction as e.g. shown for Co[23], Ni[24] or Rh[25,26] brings significant challenges for subsequent h-BN transfer. A further widely used guideline to the choice of the catalytic metal for monolayer h-BN growth is its bulk solubility of the constituent elements, whereby high B and N solubilities are considered deleterious leading to reservoir and precipitation effects upon cooling that are difficult to control[27]. Based on these selection guidelines, Cu[18,28-30] and Pt[17,31] are widely used for h-BN CVD (and also graphene for the same reasons). Coincidentally, these catalysts are not as effective for gaseous precursor dissociation[32] thus requiring high temperatures. Particularly for Cu this introduces problems due to its high vapour pressure[33] and the need for operation close to its melting point; while the use of Pt carries higher costs in particular if it cannot be continuously reused due to degradation or contamination. Yet, previous literature also indicates that the dissolution of species into the catalyst bulk can potentially offer an avenue for increased h-BN CVD control, with, for instance, the bulk reservoir effect being tunable via $NH_3$ pre-treatment for cheaper Fe catalysts[20] or interstitial C being reported to affect B and N adsorption on Co and Ni[34] surfaces. The possible interdependent effects of intentionally or unintentionally adding (e.g. via contamination of catalyst or reactor) bulk dissolved species into the catalytic growth of 2D materials are often difficult to discern and thus remain poorly understood[35]. Therefore, the dissolution of precursor species into the catalyst bulk remains more strongly associated with loss rather than gain of growth control. However, in the case of graphene CVD, careful consideration of dissolution into the catalyst bulk has allowed monolayer growth control to be extended to many high C solubility catalysts[36-38]. Recently, high temperature oxidative annealing[39,40] has emerged as a facile way to obtain large-domain graphene, with the vast majority of state-of-the-art studies on mm-sized domains exploiting this method. To date, there is no equivalent simple catalyst conditioning method that consistently results in mm-sized h-BN domains. Oxidative processing has been investigated for h-BN, but only resulted in marginal improvements[18,41].

Here, we focus on Fe catalysed CVD of monolayer h-BN. While the use of Fe is cost-effective, growth control to achieve monolayer h-BN and large h-BN domain sizes has remained a challenge due to the solubility of both B and N in Fe[20]. The transfer of h-BN away from Fe is an additional challenge due to the strong interaction[42], which is conversely why for magnetic tunnel junctions the transfer-free integration of h-BN on ferromagnetic Fe has been found so promising[43]. We employ commercial poly-crystalline Fe foils as a starting point to systematically explore the role of bulk dissolved species, and find that a simple Fe oxidation and sequential carburisation pre-growth step enable us to tailor a cost-effective, scalable CVD process to give mm-sized h-BN domains, among the largest reported to date. Building on previous literature on the use of water to facilitate graphene transfer from Cu[44], we show that a thin interfacial Fe oxide can be created which subsequently can be selectively etched to enable a much easier h-BN transfer away from Fe. We demonstrate the quality of as-grown and transferred monolayer h-BN by using it to protect graphene from damage during remote plasma assisted atomic layer deposition (PEALD) and in creating $Al_2O_3$/h-BN/Gr heterostructures. Graphene passivated with $Al_2O_3$/h-BN can withstand higher laser power densities compared to intrinsic graphene, highlighting the promise of this approach for integrated (opto-)electronic device applications.

## 2. Methods

*2.1 h-BN growth by CVD on Fe foils.*

Fe foils (Goodfellow, 0.1 mm, 99.8 % purity) were cleaned in acetone and IPA and either utilised as such or oxidised at 350 °C for 5 mins until the colour changed to brown. Such foils are not the highest purity available but are a suitable choice towards industrially applicable low-cost substrates. For instance, 99.99 % purity foils (Alfa Aesar, 0.1 mm) have similar level of surface contamination as 99.8 % foils (Goodfellow, 0.1 mm; Supplementary Figure 1a,b), since the stated purity only refers to the average bulk composition and only considers metal content. The foils were then loaded into a custom cold-wall CVD system with a SiC-coated (~40 µm) graphite heater. It should be noted that the utilisation of uncoated graphite heaters leads to uncontrolled C contamination, which must be avoided to achieve reliable process control. The system was then pumped to about $1\times10^{-5}$ mbar and filled with the required gas (Ar, $NH_3$ or $H_2$; BOC research grade) at $1\times10^{-2}$ mbar partial pressure. The temperature was then ramped to 980 °C ($T_{growth}$) or to the temperature as stated in the main text (950 °C or 1010 °C) at 50 °C/min, followed by annealing in the set gas atmosphere ($t_{anneal}$) for 20 mins. For the carburisation experiments, acetylene (BOC, research grade) was added in addition to the tested gas at a partial pressure of $3\times10^{-3}$ mbar for 5 mins ($t_{C2H2}$). Borazine (Fluorochem, >97 %) was used as the h-BN precursor and was controlled with a mass flow controller (MKS, metal seal) connected between the borazine bottle and the chamber. A borazine dose was added in addition to the tested gas, for example, 0.08 sccm for 20 mins. Different precursor flow rates ($F_{borazine}$) and growth times ($t_{growth}$) were used from 0.015 sccm to 0.54 sccm and from 5 mins to 4.5 hours respectively as described in the results section. The partial pressure of borazine in our CVD system is approximately linearly related to its flow (sccm) as follows: $(8.7\times10^{-3})\times F_{borazine}+(9\times10^{-5})$ mbar.

*2.2 h-BN transfer*

To transfer CVD h-BN from Fe substrates the samples were oxidised in an environment with a high moisture level while being heated to prevent water condensation and droplet formation that otherwise causes localized corrosion. A diagram of the apparatus is shown in Supplementary Figure S2a. To control the level of humidity in the chamber, a water bath was heated to 90 °C, while the h-BN/Fe samples were mounted on a glass cover with a secondary heater attached to the back of the glass, set to a temperature of 120 °C. After such oxidation for 24 hours, the samples were spin-coated (1000 rpm) with poly(bisphenol-A-carbonate) (PC, 5 % in chloroform), dried and floated on the surface of a 12 % solution of hydrochloric acid. After a short time, the Fe foil detached from the floating polymer support and sunk, releasing the PC/h-BN stack. The stack was then transferred to three sequential baths with deionised water and picked up with the target substrate. $SiO_2$/Si substrates (300 nm) and graphene/copper samples were used as target substrates. The samples were then dried and the PC support was dissolved in chloroform similarly to previously described methods[45].

*2.3 Characterisation.*

Scanning electron microscopy (SEM) was performed on a Carl Zeiss Gemini 300 microscope with a 5 kV accelerating voltage.

Energy-dispersive X-ray spectroscopy (EDX) and mapping was performed on a Carl Zeiss Leo scanning electron microscope with an Oxford Instruments EDX detector at 20 kV.

Time-of-flight secondary-ion mass spectrometry (ToF-SIMS) measurement were performed using a ToF-SIMS IV instrument (ION-TOF Gmbh, Germany) at a base pressure $<5\times10^{-9}$ mbar. Each depth profile was acquired by analysing a 150 µm x 150 µm surface area (128 x 128 pixels, randomly rastered) centred within a 400 µm x 400 µm sputtered region in a non-interlaced mode (alternating data acquisition and sputtering cycles). Analysis was carried out using 25 keV $Bi_3^+$ ions from a liquid metal ion gun, with a spot size less than 5 µm in spectroscopy mode, a cycle time of 100 µs, and an ion current of 0.1 pA. For sputtering, 10 keV $Cs^+$ ions with a current of 30 nA were employed. All ion beams are oriented at 45° to the sample normal. The depth was approximately estimated using the erosion rate (0.099 nm/s) calculated using the built-in software (SurfaceLab 7.0) calculator from the Fe sputter yield and sputtering beam parameters (accelerating voltage, current and raster area). To compare the carbon level in the bulk of the foil samples, the ions from the first ~5 nm of the samples were excluded, in order to remove the contribution of adventitious hydrocarbons due to ambient exposure prior to measurement. This was determined by monitoring the removal of the $CH^-$ ion signal during depth profiling.

For the analysis of representative iron oxide species, the depth-dependent $FeO^-$ (and similarly, the $C_2^-$, $FeO_2^-$ and $FeO_3^-$) ion counts were normalised by the total ion count using point-by-point normalisation to allow direct comparison between the samples. For the ToF-SIMS measurements the Fe foil samples were removed from the CVD system while the system was being vented in $N_2$ and placed in a $N_2$-filled bag. Before the ToF-SIMS measurement the samples were exposed to air for <5 mins while loading into the system and pumping to ultra-high vacuum.

Optical images of h-BN on Fe were recorded after oxidising the Fe/h-BN samples at 350 °C on a hotplate, until the colour of Fe changed. Typically, the Fe surface acquired a brown colour, but depending on the Fe grain orientation and the level of oxidation other colours could be obtained (e.g. blue). The regions protected by h-BN oxidised more slowly[43] and therefore had a different colour (e.g. light brown). Multilayer h-BN strongly protects Fe surface from oxidation and therefore resulted in a bright white colour which can be easily identified and distinguished from monolayer h-BN in the optical images. Images were recorded on a Nikon Eclipse LV150N optical microscope with software enhanced contrast.

Raman spectroscopy characterisation of monolayer h-BN was performed on an inVia Renishaw system with a 532 nm laser, 100× objective and ~10 mW power. The background of $SiO_2$/Si was subtracted, and a Lorentzian function was fitted to estimate the position and width of the $E_{2g}$ peak. Raman characterisation of graphene was performed at ~1 mW power, whereby Voigt profiles were fitted to the D, G and 2D peaks.

Selected area electron diffraction patterns were recorded on a FEI Tecnai Osiris 80-200 transmission electron microscope (TEM) at 80 kV accelerating voltage after transferring h-BN to TEM grids. For these substrates, polymethyl methacrylate (PMMA, MicroChem A4, diluted 1:2 with anisole) was dropcast onto h-BN/Fe substrates and subsequently delaminated using the electrochemical method[20]. Quantifoil TEM grids (Agar Scientific) were used to pick up the PMMA/h-BN. The samples were dried and the PMMA was dissolved in acetone.

*2.4 Graphene heterostructure fabrication with h-BN and $Al_2O_3$ and their characterisation.*

Graphene was grown by CVD on Cu foils (Alfa Aesar, 99.8 %, 25 µm) in a custom tube furnace system using previously reported methods[35,40]. Briefly, oxidised Cu foils were loaded in the CVD system and annealed at 1070 °C in 50 mbar Ar, followed by a Ar, $H_2$, $CH_4$ gas mixture for the growth stage, with flow rates of 583 sccm, 17 sccm and 0.013 sccm respectively for 5 hours to obtain continuous graphene films with mm-sized domains. The PC/h-BN stack obtained using the moisture-assisted acid release method was transferred directly to graphene/Cu, followed by etching of the Cu foil in 0.5 M ammonium persulfate solution. The resulting PC/h-BN/graphene stack was transferred to three sequential deionised water baths and picked up with the $SiO_2$/Si substrate. Similarly, for the graphene/$SiO_2$/Si samples, PC was used to coat the graphene/Cu samples. The Cu substrate was etched and the sample was rinsed in deionised water, followed by the pickup with the target substrate, drying, and the dissolution of PC in chloroform.

Plasma enhanced atomic layer deposition (PEALD) was performed in an Oxford Instruments FlexAL system. 20 nm of alumina was deposited on the graphene/$SiO_2$/Si and h-BN/graphene/$SiO_2$/Si samples at 150 °C using trimethylaluminium (TMA), oxygen plasma from a remote plasma source and water as precursors. Laser exposures to test the stability of graphene and its heterostructures were performed on an InVia Renishaw system with a 532 nm laser, ×100 objective, 100 % laser excitation (~50 mW) for 2 mins; 300 nm $SiO_2$/Si substrates were utilised. Before and after the exposure, Raman signatures of the graphene-containing samples were collected at 1 mW laser excitation.

## 3. Results and discussion

### 3.1 Pre-growth conditioning of Fe foils towards large monolayer h-BN domains

We focus on widely available, low-cost, standard purity (99.8 %), polycrystalline Fe foils to demonstrate that our process is sufficiently contamination tolerant and does not require single crystal substrate preparation or excessive metallurgical processing. To elucidate the causes of h-BN nucleation we first note an experimental observation whereby h-BN domains preferentially nucleate at the rolling striations of the as-received Fe foils (Supplementary Figure S1e). We find that the surface or near-surface contamination of Fe foils often comprises significant quantities of localised carbon (Supplementary Figure S1a-d), which we suggest leads to the observed preferential h-BN nucleation. It is therefore necessary to consider methods for catalyst surface and near-surface cleaning that would reduce the drastic effect of such embedded impurities.

A generalised CVD process flow used in this study is summarised in Figure 1a. Briefly, the Fe foils are either used pristine (after organic solvent cleaning) or purposely oxidised, then annealed ($T_{growth}$ = 980 °C) in the chosen gas atmosphere (gas 1 for the annealing and gas 2 as the carrier gas for the growth) and borazine is introduced to nucleate and grow h-BN. After the growth stage, the sample is rapidly cooled (initial cooling rate of 200 °C/min). We first investigate the annealing effects on as-received Fe foils in vacuum or typical CVD gases ($1\times10^{-2}$ mbar $NH_3$, Ar or $H_2$; gas 1 = gas 2) combined with fixed, short borazine exposure times ($t_{growth}$ = 5 mins, $t_{C2H2}$ = 0 mins; Figure 1a) in order to reveal the nucleation behaviour of h-BN domains. Figure 1b (also Supplementary Figures S3a-c) shows that such simple process in either $NH_3$, Ar, vacuum or $H_2$ results in small (< 5 µm) triangular h-BN domains with frequent multilayer patches or pyramids (light colour optically). The h-BN deposits in $H_2$ typically have higher pyramid coverage than in the other atmospheres, consistent with previous studies[20]. We conclude that such a simple process does not provide sufficient growth control.

We explore next the controlled addition of carbon to Fe foils via a short acetylene ($C_2H_2$) exposure prior to the borazine growth step (Figure 1a, gas 1 = gas 2, $t_{C2H2}$ = 5 mins, $t_{growth}$ = 5 mins; Methods). Figure 1c shows that this short carburisation step leads to a significant increase in h-BN domain size from around 5 µm (Figure 1b) to 25 µm with the $NH_3$ carrier gas and otherwise unchanged conditions. The same effect of the $C_2H_2$ step is observed for the other tested gases (Ar, vacuum and $H_2$; Supplementary Figures S3d-f respectively). The addition of carbon appears to enhance the effective h-BN layer growth rate and to increase the monolayer fraction within a given h-BN island (Figure 1c). Yet, the carburisation step does not fully suppress multilayer h-BN formation, unlike in previous reports on Co and Ni catalysed h-BN growth by molecular beam epitaxy with boron oxide and ammonia precursors[34]. We attribute this effect to the partial filling of Fe foils with C that in turn reduces the boron and nitrogen solubilities, hinted at by previous studies[46-48]. Such trends for the N solubility are presented in Supplementary Figure S4, whereby by approaching the C solubility limit in γ-Fe at 950 °C (~6.2 at%[49]), the N solubility decreases by ~30 %. A similar reduction in B solubility is expected[47], likely resulting in a significant change to the kinetic growth behaviour[36]. Carbon, boron and nitrogen are all well-known interstitial solute atoms for Fe[46-51], and therefore, C, B and N solubilities will be interdependent considering the quaternary C-B-N-Fe system. On the other hand, in the C-N-Fe system it was also shown that C and N atoms can reversibly replace each other in the bulk[51]. While the carburisation step is found to significantly affect the h-BN growth (Figure 1c), this simple process flow still results in a relatively high h-BN nucleation density and inhomogeneous growth, as highlighted by a high fraction of multilayer deposits (Figures 1c; also Supplementary Figures S3d-f).

To identify new routes of h-BN growth control we investigate oxidised Fe foils next (Methods) and optimise a suitable h-BN CVD process. Such oxidative processing is widely used to remove carbonaceous impurities in metal catalysts, for instance, to control the nucleation density of graphene for Cu catalysed CVD[40]. This change introduces new complexity, but as we will show in the following discussion, opens new pathways to h-BN growth control. When Ar is used as gas 1 and 2 in conjunction with oxidised Fe foils, we find that the dose of borazine needs to be significantly increased (by an order of magnitude) to initialise h-BN growth. We rationalise this observation by noting that h-BN growth occurs on metallic Fe, and since the Fe surface is not fully reduced in the Ar atmosphere in the given time, a higher borazine exposure is required for the full surface conversion to metallic Fe in order to initiate h-BN growth. Under these conditions, it is difficult to control the growth and as a result, very distorted, inhomogeneous domains are formed (Supplementary Figure S5a). When Ar is used as gas 1 but a reducing $NH_3$ atmosphere is used for gas 2, lower borazine pressure can be used again allowing us to obtain geometrically-shaped predominantly monolayer h-BN deposits (Supplementary Figure S5b). However, holes and damage often occur for such conditions. This observation parallels with our previous studies of bulk oxygen effects for Cu catalysed graphene growth[35], where higher defect densities were confirmed with Raman spectroscopy and increased post-growth etching in $H_2$. We find that the effect of Fe oxidation can be nullified (in terms of the h-BN material outcome) if a prolonged reducing ($NH_3$) atmosphere is used from the start (gas 1) or with the addition of a sufficient $C_2H_2$ dose with a shorter $NH_3$ annealing (Supplementary Figures S5c,d).

We systematically explored the various combinations of pre-growth Fe foil conditioning for the given process (Figure 1a) and found that a balance between bulk O and C allows advantageous h-BN growth control. Remarkably, a specific Fe foil treatment process consisting of Fe oxidation, Ar annealing (gas 1) and a short $C_2H_2$ dose prior to growth achieves large monolayer (> 100 µm) h-BN domains (Figure 1d with $NH_3$ as gas 2; also Supplementary Figures S3g-i for Ar, vacuum and $H_2$ respectively as gas 2). For all experiments, the borazine exposure time was only 5 mins, giving an average growth rate of 30 µm min$^{-1}$, one of the highest reported to date for a growth temperature below 1000 °C. The described improvement in growth outcome from Figure 1b to 1d is achieved by only targeted Fe foil conditioning prior to growth. Additionally, the h-BN does not tend to nucleate along Fe striations suggesting the removal of deleterious impurities (e.g. localised C) that otherwise would cause inhomogeneity. We find that the use of Ar as gas 1 is essential, while the choice of gas 2 is less critical. However, $H_2$ as gas 2 consistently gave worse results for monolayer h-BN coverage, likely due to its interference in the interplay between the bulk-dissolved species in Fe[20] (Supplementary Figure S3c,f,i). We converge on $NH_3$ as gas 2 to provide additional h-BN protection from oxidative gas impurities (vs. Ar) and to mitigate excessive Fe evaporation (vs. vacuum).

The developed method allows facile, rapid growth of large h-BN domains on the order of hundreds of micrometres in lateral dimensions. We also confirm good uniformity by growing h-BN domains on a 2" substrate and examining the domain size distribution (Supplementary Figure S6). The nucleation density and domain size are very consistent on this substrate and allow the growth of uniform polycrystalline films when the domains coalesce with extended growth times (Supplementary Figure S7). In the following, we focus on this narrower choice of process parameters as the framework for further growth optimisation.

### 3.2 Developing an understanding of oxidising and carburising catalyst conditioning

For clarity in the following discussion, based on previous literature[15,20,29,52,53] and our h-BN domain morphology observations above we distinguish three potential pathways to h-BN domain formation from the precursor B, N species; these are summarised schematically in Figure 2a. For a catalyst foil whose bulk is initially largely empty of dissolved B and N species, upon precursor exposure, the supply of B and N species to the surface is mediated by diffusion into the catalyst bulk. Once a sufficient supersaturation is developed at the surface, large monolayer or the first layer of h-BN multilayers can form isothermally from surface-most precursor species arising from borazine decomposition – this route is particularly important for low solubility transition metal catalysts (Figure 2b; ①; large dark monolayer h-BN domain). Isothermal growth of large additional h-BN layers underneath the first is facilitated by diffusion of species from the Fe surface and sub-surface, especially in catalysts with high precursor solubilities, akin to a h-BN layer "intercalation". While the net flow of the precursor species is still to fill the bulk of the foil with growth at the surface kinetically stabilised[36], there is also lateral precursor diffusion to grow the secondary layer which locally consumes the diffusing precursor. The concentration-driven, diffusive precursor supply via sub-surface or bulk has been directly confirmed for monolayer graphene for designed Cu-Ni alloy combinations with different C solubilities by means of the $^{13}$C isotope labelling technique[54], and with ToF-SIMS measurements of the bulk C level in Cu for graphene growth[35,55]. These h-BN layers (progressively lighter shade) look like triangular pyramids and often have the same vertex, as shown in Figure 2c. The solubilities of B and N in γ-Fe at 950 °C are around 0.11 at%[56] and 0.10 at%[46,57] (C solubility 6.2 at%[49]) respectively. These values are relatively high if put in perspective with e.g. the C solubility in Cu of around 0.03 at%[58] for graphene CVD at typical growth temperatures (~1030 °C). Even for this low precursor solubility catalytic system there is a chance of isothermal bilayer or multilayer graphene growth unless precautions are taken[35,53,55]. Another pathway for h-BN growth is precipitation from the bulk upon cooling due to the decreasing precursor solubility (Figure 2a). This process tends to result in aligned h-BN domains that are very small and thus have a noticeable dependence on the underlying grain structure of the metal catalyst as shown schematically in Figure 2a. The alignment and shape can significantly change from one Fe grain to another. Hence, such h-BN deposits can have distorted shapes dictated by the catalyst grain orientation, and are small in typical CVD processes where the growth is rapidly quenched (Figure 2a,b; ③). h-BN domains formed by precipitation do not typically occur underneath the already-grown monolayer or multilayer h-BN under with our process conditions, likely due to the energy barrier associated with decoupling the existing isothermally grown h-BN layer from the Fe surface upon rapid cooling.

It is evident from our data (Figure 1) that pre-growth processing is of utmost importance, overshadowing the typical growth parameters (precursor pressure, temperature and time). In order to elucidate the effects of bulk dissolved species and consequently the reasons for the improved growth of h-BN on the conditioned Fe foils, we utilise depth-resolved time-of-flight secondary-ion mass spectrometry (ToF-SIMS). For each sample: as-received Fe, oxidised, Ar annealed, after $C_2H_2$ exposure in $NH_3$ and after borazine exposure in $NH_3$ we analyse the negative ion intensity vs. the mass-to-charge ratio, m/z (Supplementary Figure S8a-e) and identify the $FeO^-$ and $C_2^-$ ions as prominent indicators of the dissolved oxygen and carbon. Figure 3a and b show the depth dependent profiles of these species (Methods), starting from the as-received foil and following each process step according to Figure 1a. We emphasise that the ToF-SIMS measurements refer to a post growth state after a brief ambient exposure, although precautions were taken to try to minimise air exposure (e.g. $N_2$ packaging; Methods). During CVD the profiles are expected to show a process- and time-dependent behaviour. The ToF-SIMS data shows that compared to the as-received Fe foil the oxidised Fe foil has a significantly increased intensity of $FeO^-$ species (note that the dip in the first 20 nm is due to the presence of a higher Fe oxide state, $FeO_2^-$ Supplementary Figure S9a). The $C_2^-$ counts decrease within the first ~20 nm from the surface. After Ar annealing, we find that the oxygen-related profile partially levels out and while its overall counts decrease compared to the oxidised sample (Figure 3a), the remaining level of $FeO^-$ is still very high. This behaviour is also linked to an overall significant reduction in the $C_2^-$ counts with depth (Figure 3b). In turn, following the steps of the developed CVD process, after the subsequent $C_2H_2$ exposure, the $C_2^-$ counts increase again, while the oxygen level is significantly lowered (Figure 3a, b, "$C_2H_2$ treated"). The $FeO^-$ counts of the oxygen-related profiles for the $C_2H_2$ treated and post-growth samples are close, within the noise level, however, minute O differences can be determined from other oxide species. Supplementary Figure S9b shows that the $FeO_3^-$ level in the sample after borazine exposure is slightly lower than after $C_2H_2$ treatment, suggesting that some oxygen is present during the growth and thus can potentially play an important role. Finally, the $C_2^-$ profile after growth has lower counts compared to the $C_2H_2$ treated sample, but still sufficiently high to have an effect on h-BN growth.

The ToF-SIMS data is consistent with the following interpretation of the required balance of bulk C and O. The catalyst pre-growth oxidation minimises localised C surface contamination hence giving an improved homogeneity of h-BN nucleation, which does not follow the foil striations anymore. In terms of the choice of gas 1 (Figure 1a), the utilisation of Ar is essential compared to reducing atmospheres (e.g. $NH_3$) as it allows O diffusion into the bulk such that some level of O is retained. Further C impurity removal is thus achieved due to the scavenging effect of carbon by oxygen[40], which has been exploited for centuries in steelmaking, e.g. in the Bessemer process. Too high O bulk concentration can cause etching effects (holes) in the growing h-BN, distortion of its morphology, or require too much precursor to reduce Fe and therefore results in irregular h-BN deposits. Thus, by utilising $C_2H_2$ to adjust the bulk O level[59] at a minute level and to fill the foil with C close to its solubility limit, thereby reducing the B and N solubilities, we suppress the pathway of isothermal h-BN multilayer growth while obtaining additional control of the nucleation density. Indeed, the reduction in the $B^+$ ion count between O, C-treated and untreated foils can be seen from the ToF-SIMS depth profiles and a possible reduction in the $NH^-$ ion counts is hinted as shown in Supplementary Figure S10a,b. Figure 3c schematically summarises our proposed mechanism of the coaction of O and C species as to condition the bulk of the Fe foil to supress the pathway for B, N species to feed the growth of h-BN multilayers via the bulk.

### 3.3 CVD parameter optimisation for further enlargement of h-BN domains

While the introduced simple foil conditioning procedure can easily give hundreds of micrometre monolayer h-BN domain sizes (Figure 1a,d), we now focus on the CVD growth stage optimisation to achieve larger domain sizes. Currently, h-BN domains of some hundreds of micrometres have been demonstrated[16,17,19,34,60], but reaching the next order of magnitude, beyond 1 mm, has proven difficult. Domain "stitching" in h-BN has also been proposed as an alternative way to obtain large monocrystalline domains[17,21,60], however, possible B- or N- edge termination and random occurrence of anti-parallel domain alignment mean that perfect "stitching" cannot be guaranteed[61,62]. The occurrence of such defects has been confirmed by the observation of low angle domain boundaries and overlapping domains with high resolution transmission electron and scanning tunnelling microscopies[61,63]. Hence, specifically here for h-BN, the pathway of large crystal growth via a low nucleation density approach has its merit. Additionally, the transfer methods need to be developed specifically for defect-free h-BN domains, which is more challenging and will be discussed later in the text.

Nucleation and growth require supersaturation; thus, one approach to suppress the nucleation density is to utilise lower precursor flows to approach supersaturation more slowly. The probability of new nucleation sites is thereby reduced, while the growth is still possible, albeit at a slower rate. We investigate this strategy for the fixed growth temperature ($T_{growth}$) of 980 °C (as in previous sections) by systematically lowering the precursor flow and increasing the synthesis time ($t_{growth}$) in conjunction with the developed pre-growth conditioning, so as to achieve partial coverage. We investigate $t_{growth}$ values from 5 mins to 120 mins. A selection of the optical images of the resulting large-domain (~500 µm) fully-monolayer h-BN deposits are shown in Supplementary Figures S11a,b. There is a rapid decrease in the borazine flow rate required to obtain h-BN deposits vs. time, which becomes less pronounced for longer times as shown in Figure 4a. Our study here links to previous studies of the incubation time for graphene nucleation, for example on Ni[36] – a strongly interacting substrate with substantial C solubility. It was determined that the impingement flux of the precursor is inversely proportional to the incubation time squared[36]. Interestingly, if we apply such a fit to our data (Figure 4a, red curve), the behaviour does not match well for short times. It is as if the necessary flow of the precursor for short times is higher than predicted by the theory. This behaviour can be again rationalised by the O etching effect due to its diffusion from the bulk of the Fe foil in our process. The additional precursor flow is necessary to compensate for the removal of O. It is also evident that the effect becomes less pronounced for longer synthesis times, suggesting that O gradually depletes and has less of an effect. Figure 4b shows that there is a clear correlation between increased h-BN domain sizes and prolonged growth times, which is most pronounced for the growth time up to one hour where the average domain size increased from around 200 µm (5 mins) to 400 µm (60 mins); with maximum domain sizes of up to 600 µm (Supplementary Figure S11a,b). Beyond this time there is only a moderate increase, suggesting that the bulk Fe foil composition is no longer suitable for the suppression of new, secondary nucleation. For extended experiments we also observe the appearance of h-BN multilayer patches which is also attributable to the time-dependent change in the foil composition likely caused by the reduction of O and C content.

Another well-known approach to achieve lower nucleation density for 2D material growth is to utilise higher growth temperatures[17,54,64]. Figure 5a shows the results for a CVD process temperature (Figure 1a) of 1010 °C, whereby we also slightly lowered the borazine flow rate. The h-BN domain size increases significantly with domain dimensions reaching 1 mm even for a relatively short $t_{growth}$ of 45 mins. However, at the same time, these h-BN domains show a high multilayer fraction, as highlighted in the optical images by the lighter deposit in the h-BN domain centre. This observation is consistent with our model introduced above, whereby the diffusion of O and C is accelerated by temperature, thus the species balance is rapidly changed, allowing the increased lateral pathway of the B- and N- species through the bulk of the catalyst, and hence the formation of multilayer h-BN growth sites underneath the existing first h-BN layer. With this understanding, instead, we lowered the CVD process temperature to 950 °C, whereby the suitable bulk composition of the Fe foil persists for longer, in favour of longer synthesis times ($t_{growth}$ = 4.5 hours) to supress the h-BN nucleation density using very low precursor flows ($F_{borazine}$ = 0.01 sccm). Figure 6b shows that under these conditions fully monolayer h-BN domains form with dimensions up to 1.1 mm.

Notably, this optimised, lower temperature of 950 °C is well within standard quartzware operational regimes (< 1100 °C) and is straightforward to implement in common CVD reactors. We here focused on a simple CVD process flow (Figure 1a) which can possibly be further optimised to achieve faster processing times by utilising, for instance, a two-step procedure, whereby a higher precursor "pulse" would rapidly trigger supersaturation and h-BN nucleation, followed by a lower precursor dose for domain expansion[17].

### 3.4 Catalyst interface engineering for improved h-BN transferability

For most applications scalable and clean transfer of the 2D layer away from the catalyst surface is required. This is a significant challenge particularly as many catalytic metals that are very good growth substrates, such as Ni, Co and Fe, have a strong interaction[22] with h-BN and thus do not allow easy h-BN layer removal. Such substrates typically require full wet etching of the metal foil for h-BN transfer[23,24]. Metal foil etching is well known to add contamination to the transferred 2D layers[65]. This contamination does not just come from the catalyst metal itself but also from the release of species/contamination trapped in the bulk of the foil, e.g. Si-containing impurities[65,66]. For our carburised Fe foils, we clearly observe that a porous "sponge" of carbon is formed after the foil is dissolved (Supplementary Figure S12a,b). This observation is an indirect confirmation of the level of C bulk dissolution, consistent with the growth model discussed above. Previous work[34] on h-BN growth on carburised Co and Ni utilised full foil etching in order to release the h-BN layer, however, the contamination issue was not discussed. Aside from environmental and economic considerations, clearly full foil etching is not an ideal transfer method, and especially not when bulk dissolution of species is used to tune growth, as we introduced above. Two alternative transfer techniques exists: bubbling transfer[67] and dry peeling[17], but both often result in macroscopic damage and tearing, especially for the strongly-interacting catalysts. Depending on which metal catalyst is used, there are currently trade-offs for each method; specifically between scalability, cleanliness, cost and levels of defects introduced into the 2D layer[68]. It should also be emphasised that the larger in size and the better in crystal quality the 2D material domains are, the harder transfer may get with these non-destructive techniques[17]. Our previous work[42] highlighted the challenges of h-BN transfer away from Fe, hence we seek here an improved transfer process matched to our catalyst and process flow (Figure 1a).

The susceptibility of Fe to corrosion is well known, especially in moist environments due to the formation of non-passivating oxides[69]. As much as this corrosion behaviour is undesirable, e.g. in metallurgy, it turns out to be very useful to exploit for facile h-BN transfer. While previous works[43] with in-situ X-ray photoelectron spectroscopy showed that a h-BN layer can temporarily protect Fe from oxidation in ambient conditions for around a week, we find that in humid environments h-BN does not provide a good protection. The possibility of moisture penetration through the rust layer potentially allows to overcome the problem of the strong Fe/h-BN interaction. Figure 6 shows the schematic rationale and results of this transfer approach. The key point as highlighted in Figure 6a is that the Fe surface oxide is not self-passivating in warm humid environments, i.e. the propagation of the surface oxide can partially proceed via the already-formed oxide. The processing steps are described in the Methods section whereby a heated, controlled humidity chamber was utilised (Supplementary Figure S2a). This process indeed occurs as shown by the change in contrast (dark to light) in SEM (Figure 6b,c) and in colour (light grey to brown) optically (Supplementary Figure S2b). Catalyst-water oxidation was previously utilised for the Cu-graphene system, albeit by submerging in water, rather than humidity control[44]. After the oxidation and coating the Fe/h-BN material with a temporary polymer support, the sample is floated on the surface of an HCl solution (Methods). Fe oxides are etched much quicker (~minutes) than metallic Fe, therefore the surface oxide is rapidly and selectively removed (Figure 6d). The h-BN/polymer layer spontaneously detaches (Figure 6e,f), as demonstrated here for 2" substrates, and can be transferred to the target substrate. Figure 7 shows optical, Raman and selected area electron diffraction (SAED) characterisation of as-transferred h-BN monolayer domains.

After transfer to 300 nm $SiO_2$/Si substrates there is a notable h-BN Raman peak with position of about 1371 $cm^{-1}$ and width of 15 $cm^{-1}$ (13-17 $cm^{-1}$ range). The former value is blue-shifted from the bulk value (1365 $cm^{-1}$) and is typical for monolayer h-BN[17,70], while the low width value is comparable to exfoliated monolayer h-BN[70] (typically 11-18 $cm^{-1}$) and the highest quality CVD monolayer h-BN[17,34] (13-16 $cm^{-1}$). We perform selected area electron diffraction (SAED) after transferring monolayer h-BN films consisting of 100-300 µm domains to TEM grids (Methods). Figure 7c shows SAED patterns spanning a distance of 250 µm which do not show any rotation, confirming the crystallinity of such h-BN domains.

### 3.5 Monolayer h-BN as graphene protection in integrated manufacturing

In order to highlight the quality and usability of as-grown h-BN monolayer films and the introduced transfer process, we here look at its use as an interfacial buffer layer between a monolayer graphene film and a commonly deposited dielectric, $Al_2O_3$. We also adapt our methods to achieve a polymer-impurity free interface between h-BN and graphene reported by previous studies[71,72]. Here, the PC/h-BN stack is directly transferred to graphene/Cu and the resulting PC/h-BN/graphene stack is delaminated (Methods). After transferring the h-BN/graphene heterostructure onto $SiO_2$/Si substrates we then employ remote plasma enhanced atomic layer deposition (PEALD) to deposit a 20 nm thick $Al_2O_3$ layer. We investigate the effect of this deposition on graphene and h-BN/graphene as shown schematically in the insets of Figure 8a,b. Compared to thermal ALD, PEALD allows deposition at lower temperatures, and potentially on plastic substrates, with wider precursor chemistries. Such processing thus opens integration routes with widely available plastic substrates for flexible, transparent electronics, where the 2D materials are utilised as active layers. The h-BN layer here is crucial to protect the graphene layer from any damage during the PEALD. Without h-BN the monolayer graphene region (scan size 50 µm × 50 µm, Methods) shows a significant increase in the Raman D:G ratio, from 2.5 % to 10 %, after PEALD at the given conditions (Figure 8a). While the intrinsic defect density in our monolayer CVD graphene is very low, the damaging effect of PEALD could potentially be different if the intrinsic defect density is lowered even further, for example as in mechanically exfoliated graphene pieces[73]. Remarkably, a single layer of h-BN is able to protect graphene from such remote plasma damage, with no measurable change in the D:G ratio before and after PEALD as seen from the histograms in Figure 8b. Also, the FWHM of the 2D peak or other graphene characteristics were not significantly affected (Supplementary Figures S13a-f), thus allowing us to create a high quality $Al_2O_3$/h-BN/graphene heterostructure.

Interestingly, we find that such $Al_2O_3$/h-BN/graphene heterostructures exhibit significantly improved stability under laser excitation, compared to just unencapsulated 2D layers. Figure 8c compares pristine graphene, h-BN/graphene and $Al_2O_3$/h-BN/graphene all on $SiO_2$/Si (300 nm) substrates after exposure to very high laser power densities. A 20 mW laser, 532 nm laser was utilised with a 100× objective with an approximate power density of ~0.5 $W\mu m^{-2}$ for 2 mins (Methods). Pristine high-quality monolayer graphene (~1 mm domain size, Methods) is predominantly destroyed, as highlighted by the very high and broad D and G peaks with a ratio of 70% after the laser exposure (Figure 8c). The 2D peak is also significantly broadened. Monolayer-h-BN-protected graphene performed better, but still, a very high D peak appeared (D:G ratio of 30 %), and broadening was observed of the G and 2D peaks. In contrast, the $Al_2O_3$/h-BN/graphene sample did not show any significant degradation, with only a minor increase in the D:G ratio (from 2.5 % to 3 %), while the widths of the 2D and G peaks remained unchanged. We hypothesise that the thicker alumina layer is an excellent barrier to atmospheric gases, which supresses thermal- or photo-activated chemical damage to graphene under such high laser power densities. It is thus evident that the h-BN monolayer enables defect-free graphene passivation with PEALD, which potentially can be expanded to other oxygen plasma-sensitive materials, such as organic films or perovskites. Additionally, graphene passivated with alumina and h-BN has enhanced resistance to high laser power densities, which is required in applications like mode-locked lasers, optical modulators or high power photodetectors.

## 4. Conclusions

In this work we present a route for monolayer h-BN utilisation towards applications – starting from the growth optimisation and developing a tailored transfer process for integration with graphene to enhance its performance. Facile conditioning of widely available iron foils was developed based on the oxygen-carbon coaction to reach hundreds of micrometre h-BN domain sizes, which was further improved by optimisation of the synthesis parameters to reach domains sizes above a millimetre. A simple gas conditioning involving oxidative annealing in argon, and a reducing step with acetylene is employed; in direct parallel to the well-established methods for mm-sized graphene growth on Cu. A targeted transfer method was developed for the Fe foils by controlled humidity oxidation and acid release of h-BN, demonstrated on large-area substrates. We show that the 2D material transfer process cannot be viewed as generic but must be tweaked for a particular catalyst or even specifically to the processing of the catalyst. Iron is particularly suited for such a process as it does not form passivating oxides and in this respect is superior to other strong catalysts such as Ni or Co. Finally, we show a highly beneficial utilisation of monolayer h-BN films as an enabling protective layer for plasma enhanced atomic layer deposition, which in turn gives improved stability under high power laser excitation paving the way for low-defect (opto-)electronic devices based on 2D materials.

## Acknowledgements

V.B., M.O., D.N. and S.H. acknowledge funding from the European Union's Horizon 2020 research and innovation program under grant agreement No number 785219. B.C. acknowledges funding from the European Union's Horizon 2020 research and innovation program under grant agreement No number 796388. J.A.-W. acknowledges the support of his Research Fellowship from the Royal Commission for the Exhibition of 1851. R.S.W. acknowledges a EU Marie Skłodowska-Curie Individual Fellowship (Global) under grant ARTIST (No. 656870) from the European Union's Horizon 2020 research and innovation programme. Y. F., V.-P. V.-R., O.B. and S.H. acknowledge funding from EPSRC (EP/P005152/1, and Doctoral Training Award EP/M508007/1). V.-P. V.-R. further acknowledges support from NPL. B.B. and A.J.P. acknowledge funding from the U.K. Department of Business, Energy and Industrial Strategy (NPL Project Number 121452).

## ORCID iDs

Vitaliy Babenko 0000-0001-5372-6487
Barry Brennan 0000-0002-5754-4100
Barbara Canto 0000-0001-5885-9852
Robert Weatherup 0000-0002-3993-9045
Stephan Hofmann 0000-0001-6375-1459

Figures

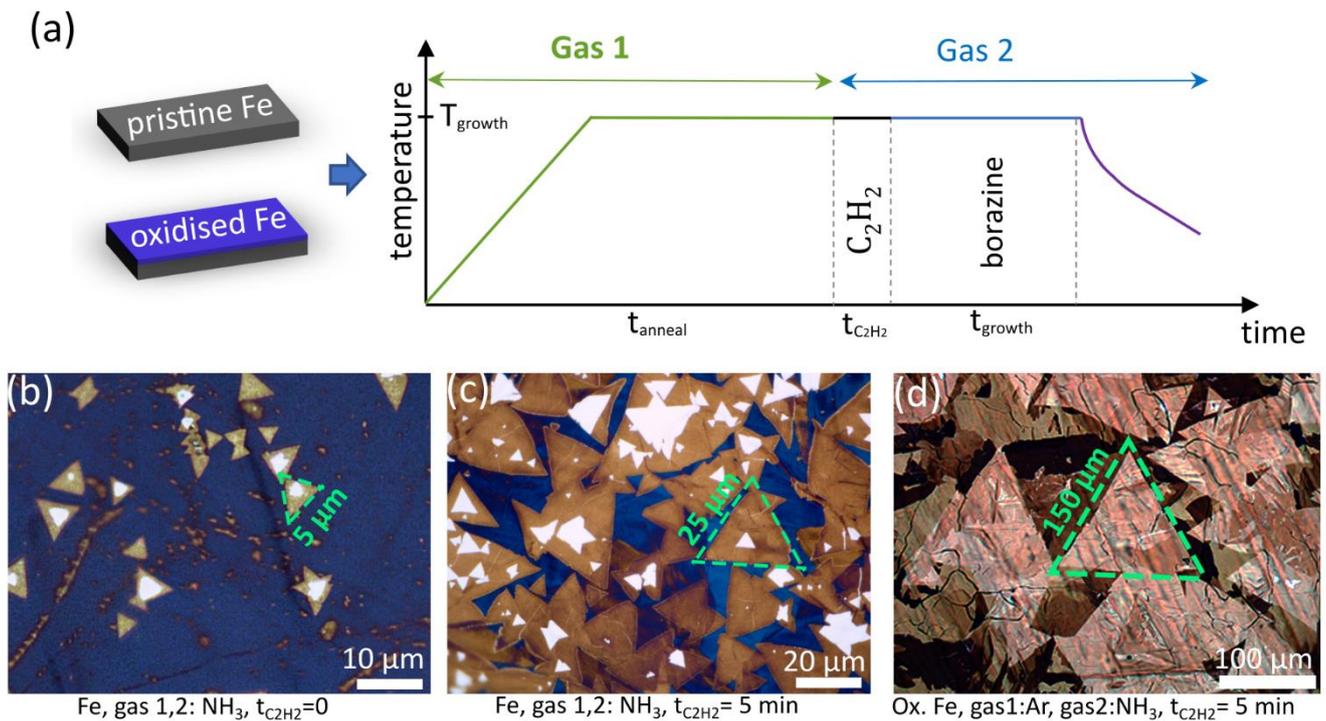

**Figure 1.** The effect of different catalyst processing steps on the h-BN domain morphology. The growth time ($t_{growth}$) and growth temperature ($T_{growth}$) for (b-d) are 5 mins and 980 °C respectively. (a) The CVD process diagram showing that the Fe foil can be pristine or oxidised; can be annealed in different gas atmospheres and can be carburised with acetylene before h-BN growth with borazine. Optical images of the h-BN deposits after post-growth Fe oxidation are shown. Fe grains can be seen as background of different colours; lighter triangular shapes are monolayer h-BN and white deposits are multilayer pyramid-like h-BN deposits. (b) A $NH_3$ based simple growth process on pristine Fe showing small triangular pyramid-like domains; additional growth results on pristine Fe foils for Ar, vacuum and $H_2$ are shown in Supplementary Figure S3a-c (c) A $NH_3$ based process with $C_2H_2$ addition before h-BN synthesis on pristine Fe foils, resulting in larger lateral monolayer deposits. Additional experiments for the Ar, vacuum and $H_2$ based processes with $C_2H_2$ are shown in Supplementary Figure S3d-f. (d) Synthesis of large, fully monolayer h-BN domains on oxidised Fe foils with $C_2H_2$ addition based on Ar annealing (gas 1) and $NH_3$ carrier gas (gas 2) during growth. Additional experiments on foils conditioned in such a way for Ar, vacuum and $H_2$ are shown in Supplementary Figure S3g-i.

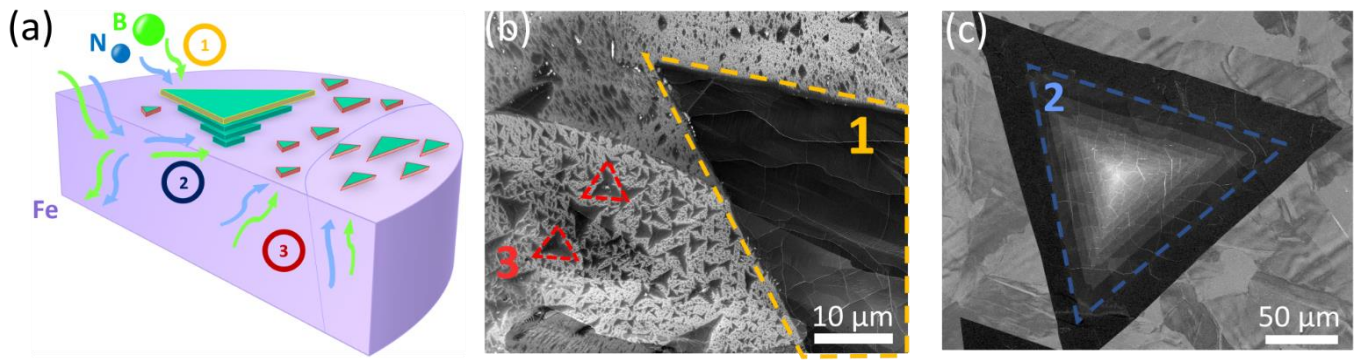

**Figure 2**. An outline of h-BN growth routes on iron. (a) A schematic diagram of the paths of B, N- species to h-BN growth sites: (1) isothermal growth of the top domain from surface species, (2) isothermal growth of large secondary layers via layer "intercalation" with species diffusing through the bulk/near-surface resulting in an inverted pyramidal shape; (3) growth by precipitation from the bulk on cooling resulting in small h-BN domains that are often aligned or have disturbed but consistent shape within a specific Fe grain. Such domains do not appear underneath the isothermally-grown layers or full coverage films. (b-c) SEM images of a monolayer regions and a multilayer pyramid illustrating exemplar domains grown via paths 1-3.

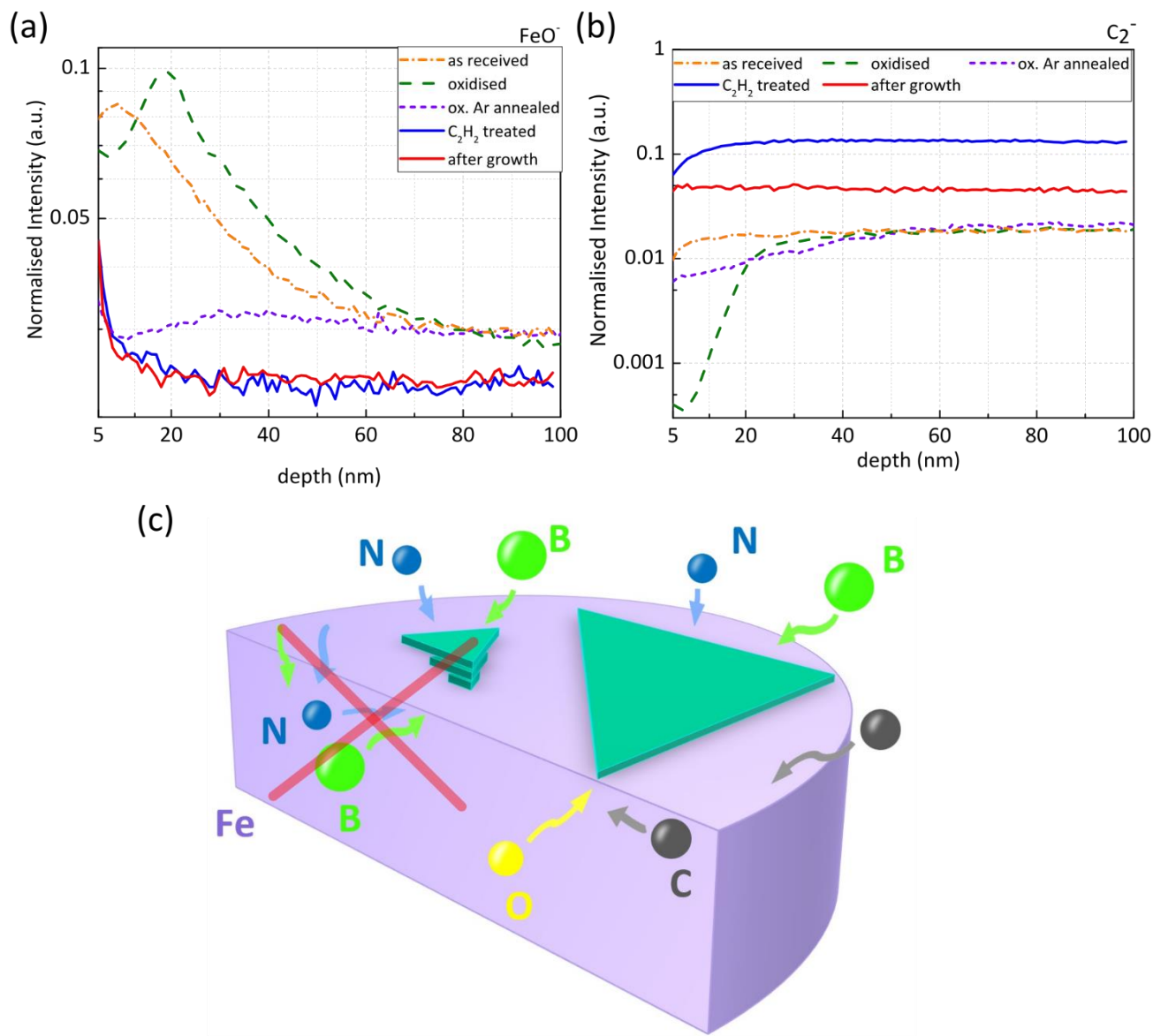

**Figure 3**. Fe foil composition analysis close to the surface for h-BN growth. **(a-b)** ToF-SIMS measurements of oxygen and carbon levels in the Fe foil, represented by FeO$^-$ and C$_2^-$ intensities normalised to total ion counts (Methods), at different stages of the synthesis process: (1) as-received foil contains some O and C; (2) Fe foil oxidation leads to the removal of most of the C within the first 20 nm of the surface and a significant increase in surface O; (3) Ar annealing of the oxidised sample causes a reduction in the O intensity, while decreasing the level of C in the foil; (4) after C$_2$H$_2$ dosing, there is a significant decrease in the O level and an increase in the bulk C level; (5) after the growth of h-BN domains a small further decrease in the already low iron oxide level can be measured (Supplementary Figure S9), and additionally the level of C is decreased, suggesting its role in the h-BN growth in pre-filling the bulk. **(c)** A schematic diagram showing that the isothermal secondary layer growth process is suppressed with the presence of oxygen and carbon in the bulk of the Fe foil leading to monolayer-only growth and suppressed nucleation density.

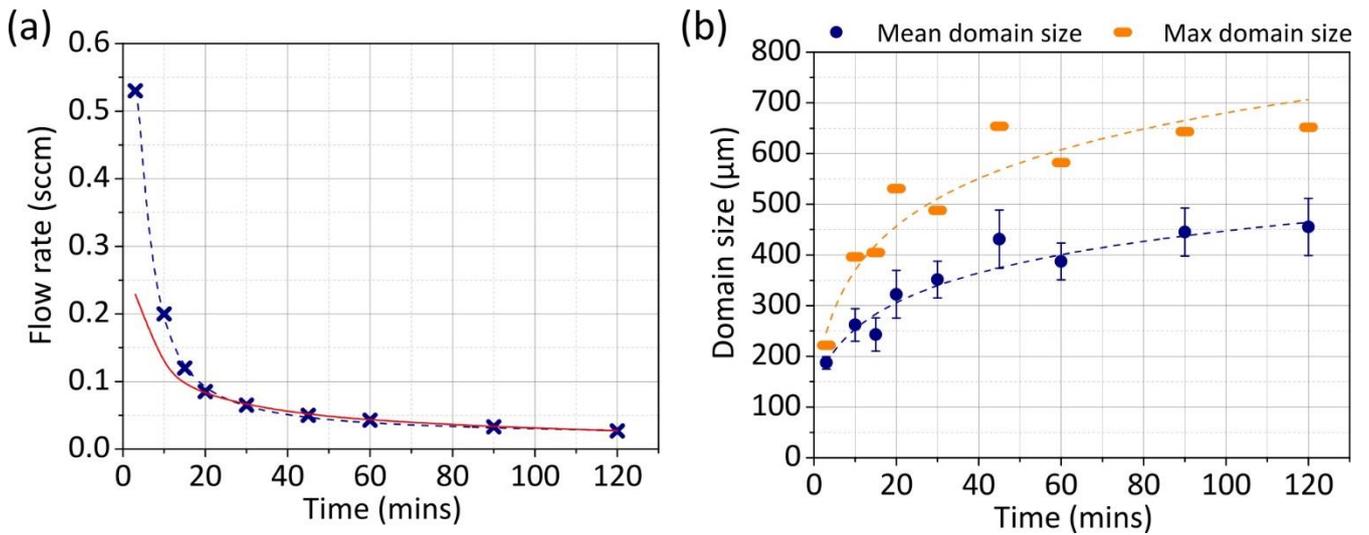

**Figure 4**. h-BN growth optimisation for combinations of precursor flow and growth times (t$_{growth}$) for a fixed temperature of T$_{growth}$ = 980 °C. (a) Borazine flow rate needed for a specific growth time to obtain separated h-BN domains; showing a rapid decrease. The red solid line is the inverse quadratic fit expected for the pristine metal catalyst to reach supersaturation, fitted to the points for t$_{growth}$ > 60 mins. The dashed line is a guide to the eye. (b) Triangular h-BN domain sizes (mean, maximum and 1 s.d. error bars are plotted) corresponding to the t$_{growth}$ times from (a).

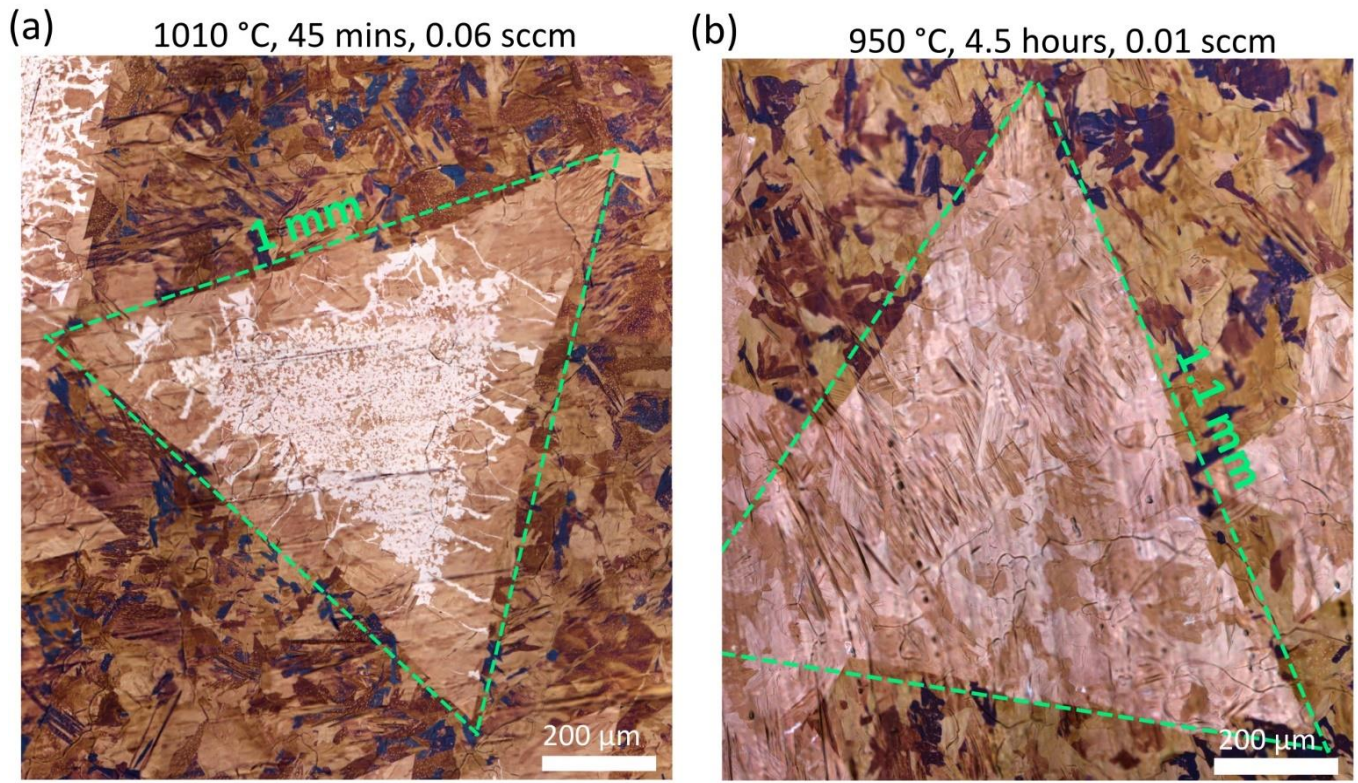

**Figure 5**. Optical images of ultra-large h-BN domains after Fe foil oxidation. (a) A h-BN domain grown at increased temperature (1010 °C) showing a large monolayer domain and patches of multilayer (white). (b) A h-BN domain grown at a reduced temperature (950 °C), but increased time showing a fully monolayer domain.

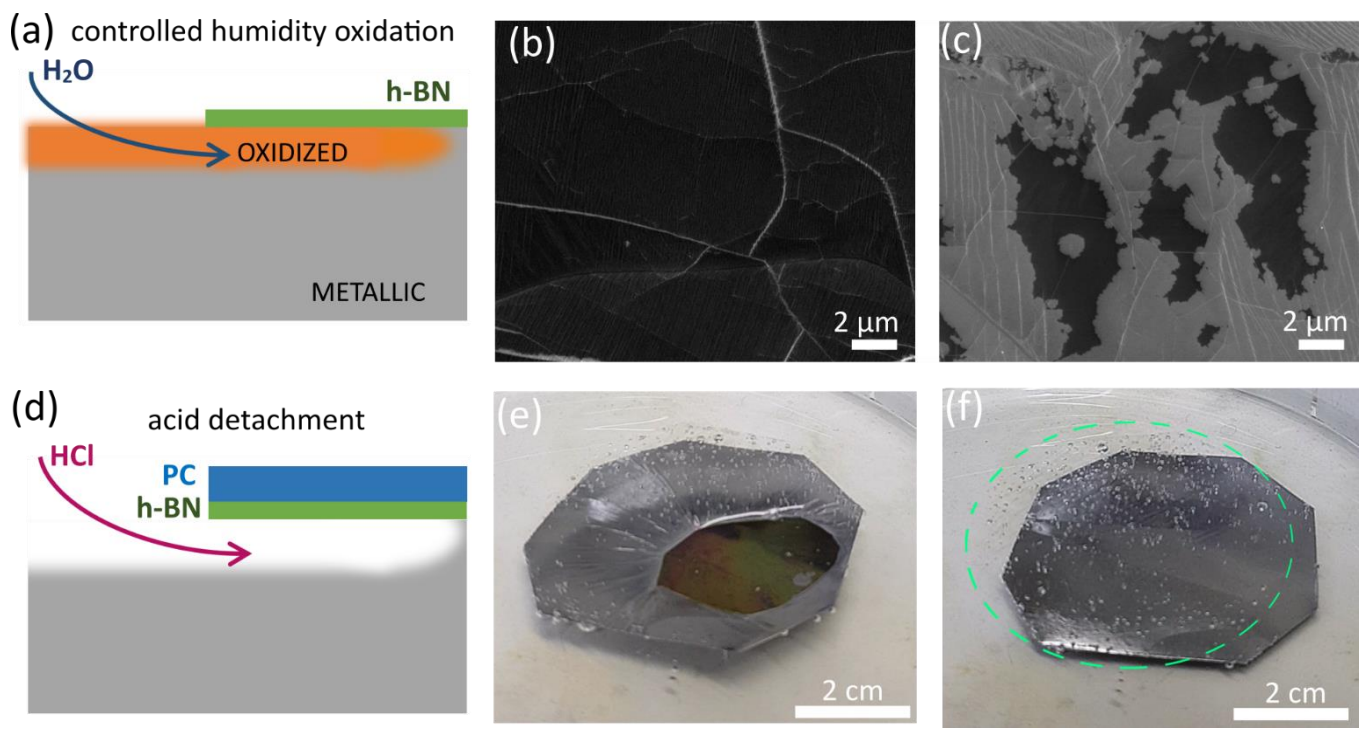

**Figure 6**. Tailored transfer of h-BN from Fe foils. (a) A diagram of oxidation of Fe underneath h-BN films via non-passivating iron oxides in humid environments. SEM images of the surface using the InLens detector showing (b) unoxidised foils with h-BN immediately after growth; (c) partially oxidized surface with oxidation in a controlled-humidity apparatus (Supplementary Figure S2a). An optical image also showing a colour change is presented in Supplementary Figure S2b. (d) A diagram of HCl etching of the thin surface iron oxides leading to PC/h-BN delamination. Optical images of the PC/h-BN/Fe samples (e) during and (f) after HCl delamination.

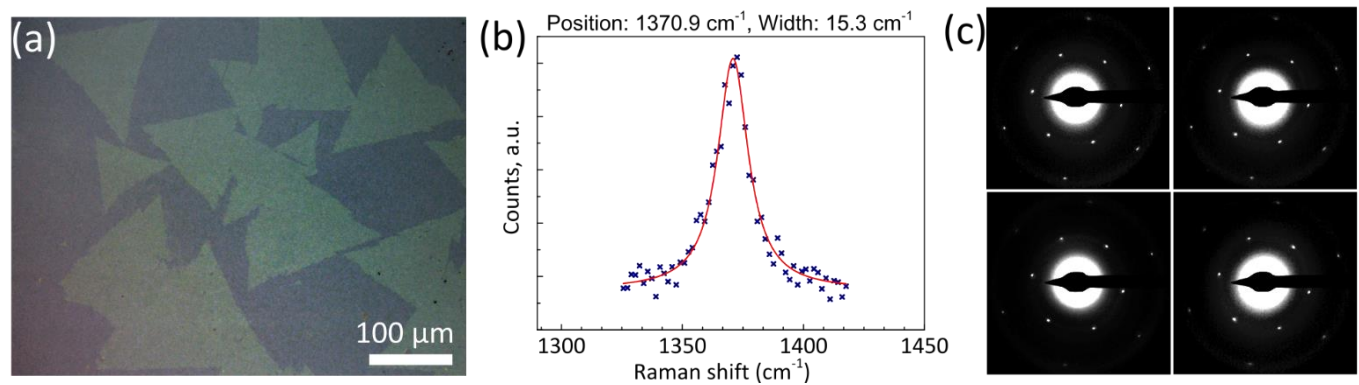

**Figure 7**. h-BN material characterisation. (a) An optical image showing exemplar transferred monolayer h-BN domains, (b) Raman signal after subtracting the SiO$_2$/Si background showing the E$_{2g}$ peak position at 1371 cm$^{-1}$ with FWHM of around 15 cm$^{-1}$, (c) SAED TEM analysis of a h-BN domain spanning a distance of 250 μm without changes in the diffraction pattern.

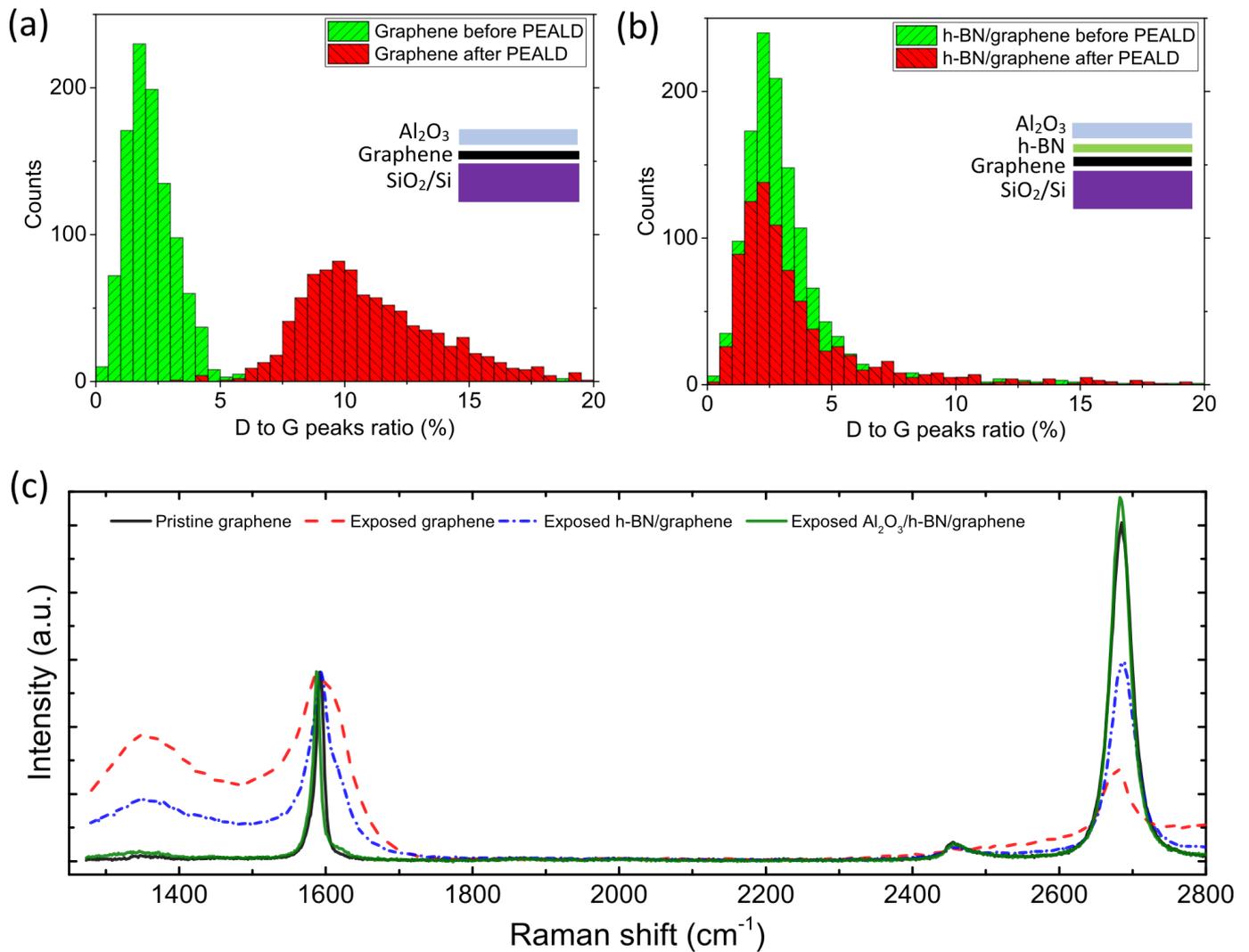

**Figure 8**. Raman spectroscopy characterisation of graphene with h-BN and PEALD alumina. (a) The D:G ratio histograms of graphene transferred to Si/SiO$_2$ substrates before and after PEALD of alumina showing a significant increase in the ratio and therefore the defect density in graphene due to the plasma induced damage. (b) The D:G ratio histograms of h-BN/Gr before and after PEALD of alumina showing no change in the ratio and therefore unchanged defect density. (c) Raman spectra of graphene exposed to high laser power densities for 2 mins (100× objective) showing that pristine graphene is significantly damaged after the exposure. While h-BN provided some protection, a significant D peak emerged and the 2D and G peaks broadened. Notably, alumina encapsulated graphene, enabled by h-BN protection, did not show a significant change in the spectrum after such laser exposure.